# A transparent waveguide chip for versatile TIRF-based microscopy and nanoscopy

Anish Priyadarshi[‡a,b*], Firehun Tsige Dullo[‡a], Deanna L. Wolfson[a], Azeem Ahmad[a], Nikhil Jayakumar[a], Vishesh Dubey[a], Jean-Claude Tinguely[a], Balpreet Singh Ahluwalia[a*] and Ganapathy Senthil Murugan[b*]

[a] Department of Physics and Technology, UiT The Arctic University of Norway, 9037 Tromsø, Norway

[b] Optoelectronics Research Centre, University of Southampton, Southampton SO17 1BJ, United Kingdom

‡ contributed equally to this work.

**Corresponding emails:** anish.priyadarshi@uit.no, balpreet.singh.ahluwalia@uit.no , smg@orc.soton.ac.uk

**Abstract**

Total internal reflection fluorescence microscopy (TIRF) has enabled low-background, live-cell friendly imaging of cell surfaces and other thin samples thanks to the shallow penetration of the evanescent light field into the sample. The implementation of TIRF on optical waveguide chips (c-TIRF) has overcome historical limitations on the magnification and field of view (FOV) compared to lens-based TIRF, and further allows the light to be guided in complicated patterns that can be used for advanced imaging techniques or selective stimulation of the sample. However, the opacity of the chips themselves has thus far precluded their use on inverted microscopes and complicated sample preparation and handling. In this work, we introduce a new platform for c-TIRF imaging based on a transparent substrate, which is fully compatible with sample handling and imaging procedures commonly used with a standard #1.5 glass coverslip, and is fabricated using standard complementary metal-oxide-semiconductor (CMOS) techniques, which can easily be scaled up for mass production. We demonstrate its performance on synthetic and biological samples using both upright and inverted microscopes, and show how it can be extended to super-resolution applications, achieving a resolution of 116 nm using super resolution radial fluctuations (SRRF). These new chips retain the scalable FOV of opaque chip-based TIRF and the high axial resolution of TIRF, and have the versatility to be used with many different objective lenses, microscopy methods, and handling techniques. We thus see c-TIRF as a technology primed for widespread adoption, increasing both TIRF's accessibility to users and the range of applications that can benefit from it.

## 1. Introduction

A common trade-off in microscopy is between resolution, speed, and photodamage or photobleaching. Total internal reflection fluorescence (TIRF) microscopy, however, is a rare example of a method which improves all three simultaneously, compared to confocal microscopy. In TIRF, only the bottom ~100 nm of the sample is excited [1-4], which improves the axial resolution, eliminates out-of-focus light, and protects the bulk of the sample from photodamage. Additionally, because TIRF is a widefield technique, acquisition of a full image can occur within milliseconds. TIRF's ability to focus exclusively on a thin layer at the surface of the cell have made it an excellent tool for studying, among others, the dynamics of focal adhesions[5], the inner workings of endocytosis[6], the kinetics of cell surface receptors[7], and docking of synaptic vesicles with neurons[8].

The enabling mechanism of TIRF is the generation of a thin, exponentially decaying layer of light at a surface, called the evanescent field. When light is directed into an interface between media with a high index of refraction contrast (HIC) at a sufficiently high angle, the light is totally reflected within that interface; while the light itself does not escape from the high index material, an evanescent field is generated along the surface that the light travels[2,9]. The angle necessary for TIRF has traditionally been achieved using a high numerical aperture (NA) objective lens[1,10,11], through a prism[1,4,11], through the use of grating couplers[12-15] and more recently by coupling into the side facet of optical waveguide chips[16-21]. These photonic chips are fabricated using technology similar to computer chips, and thus have the potential to be mass produced at low cost. While objective lens-based TIRF is restricted to using a high NA lens, thus limiting its field of view (FOV), the evanescent field generated by waveguides is independent from the imaging pathway, enabling them to be used with any imaging objective on a standard microscope[12,18,20]. Furthermore, waveguide chips have been extended to super-resolution modalities, including single molecule localization microscopy (SMLM)[18,21], entropy based super-resolution imaging (ESI)[19], and structured illumination microscopy (SIM)[21]. In addition to achieving a sub-diffraction localization precision of 72 nm, chip-based SMLM was able to do this with an unprecedented FOV of 0.5x0.5 mm$^2$, approximately 100 times larger than objective lens based techniques[18]. Similarly, chip-based TIRF-SIM has surpassed the 2X resolution enhancement of conventional SIM, achieving a 2.4X resolution enhancement due additional benefits from the fringe pattern being generated in a high index material instead of free space[21].

Thus far, most chip-based microscopy has been performed on waveguides fabricated on top of opaque substrates[12,16-18,20,21]. With the sample sitting on top of the waveguide, the chips can therefore only be used with an upright microscope (Fig 1a) because the emitted light cannot pass through the opaque substrate to the objective of an inverted microscope. On an upright microscope, however, the light must travel from the bottom of the sample where it is emitted, through the rest of the sample (which may be complex and highly scattering) and its surrounding media, and typically through a glass coverslip to the objective. This translates to aberrations in the image, not because of inferior optics but because of the scattering of light from the sample itself and from index mismatches from the medium in which cells are kept (Fig. 1a). This problem is even more pronounced for super-resolution techniques, for example in SMLM, where aberrations limit both the localization precision of the single molecule and the resulting resolution enhancements, and in SIM where these aberrations can result in additional reconstruction artefacts.

Additionally, opaque chips have several practical handling difficulties, particularly with their use on an upright microscope[11,22]. Typically, a thin PDMS frame is placed on the chip around the specimen to contain the media, and a glass coverslip seals the chamber from the top during imaging; while this limits evaporation from the sample, it precludes the addition of reagents or the use of micromanipulation tools during imaging, the low volume can be problematic for live cell imaging, and the chamber often induces tilt in the coverslip, leading to additional image aberrations. Unless the objective is designed for water, the media in the chamber induces yet more aberrations from index mismatch. Furthermore, the opacity of the chip makes it impossible to check on the sample using a basic brightfield or phase contrast microscope, so the confluency and health of cells cannot be checked prior to labelling and imaging.

Thus, having a transparent chip platform (Fig. 1b) would be highly beneficial, so that the sample can be more easily handled and inspected, remain open on the top for sample treatment during the experiment, and be used with a broader variety of objective lenses and microscopes. For the design of a transparent waveguide chip compatible with these goals, the key is to match its properties as closely as possible to a standard #1.5 glass coverslip[10,23]. Previous coverslips which used a grating to couple in light for TIRF[12-14] used traditional glass coverslips for their substrate, which are incompatible with the annealing process necessary for high-quality waveguides, therefore limiting the intensity that could be achieved during imaging and eliminating the ability to guide light along specified paths for use with e.g. fluorescence recovery after photobleaching (FRAP) or photoactivation experiments. Compatibility can be recovered by choosing a substrate other than standard glass, although compatibility with standard sample preparation and imaging processes must be maintained. Choice of a different substrate can also help ensure HIC, which is needed to ensure low propagation losses in the waveguide, resulting in a larger proportion of the chip having sufficiently high intensity for imaging[24].

Keeping within these guidelines will result in a much more versatile and user-friendly tool thereby increasing the likelihood of adoption by the broader biological imaging community. Commercially-available Zeptosens waveguide slides, which are based on thick transparent glass slides (and are therefore incompatible with high NA, short working distance objectives) have already shown that there is a strong market for user-friendly TIRF technologies[12,13], thus further expansion into high-performance, coverslip-compatible waveguide TIRF technologies is likely to still be more beneficial. In this work, we have developed an optical waveguide platform based on HIC waveguide materials: tantalum pentoxide on a thin, transparent fused silica substrate, which mimics standard glass coverslips and is compatible with standard complementary metal oxide semiconductor (CMOS) mass-production processes. We show that it can be used for TIRF imaging on both upright and inverted microscopes, and demonstrate this using both fixed and living samples. Furthermore, we show that this chip platform is compatible with fluctuation-based super-resolution imaging, in particular super-resolution radial fluctuations (SRRF)[25].

**2. Design of transparent waveguide chip**

To ensure the highest compatibility with a wide variety of microscopes, the waveguide chips should closely match the properties of a conventional glass coverslip, including the transparency, thickness, and index of refraction (RI). Additionally, the material must be compatible with the waveguide fabrication processes, have low surface roughness, and ideally be relatively inexpensive and readily available. The borosilicate glass used for traditional coverslips is unfortunately incompatible with the higher-temperature fabrication steps (such as annealing), and has a somewhat high propagation loss[26]. While sapphire is both transparent and compatible with high-temperature processes, it is expensive and has an RI significantly different from that of coverslip glass ($n_{sapphire}$ = 1.70[27] compared to $n_{glass}$ = 1.51 @ 633 nm[23]). Fused silica (as opposed to oxidized silica) is transparent at visible wavelengths, and compatible with annealing. Additionally, its RI ($n_{fusedsilica}$=1.46) is relatively close to that of glass, and its propagation losses are lower than those in glass[28]. Thus, fused silica was chosen as the waveguide substrate, and a thickness of 180 μm was chosen for compatibility with #1.5 borosilicate coverslips. Spherical aberrations resulting from a difference in the optical path length through the fused silica substrate compared to a conventional coverslip, due to their different RIs, can be compensated for either through chemical mechanical polishing[29] to reduce the thickness of the fused silica, by adjusting the correction collar found on some higher-end microscope objectives[30], or, for oil-immersion objective lenses, by changing the RI of the immersion oil.

c-TIRF applications benefit from a strong evanescent field as it produces increased fluorophore emission and enables blinking (on-off switching) for super-resolution applications. Thin waveguide geometries (e.g. ~200 nm thickness) made from HIC materials such as $Si_3N_4$ and $Ta_2O_5$ produce strong evanescent fields,[18,24] making them good choices for the transparent waveguide platform. $Ta_2O_5$ has a higher RI than $Si_3N_4$ (>2.1[31] compared to 2.007[32] at 632 nm), and as it was recently reported that $Si_3N_4$ has significantly higher background signal, especially for the 400-500 nm wavelength region[33], we choose $Ta_2O_5$ for the c-TIRF waveguides.

Fabrication of these waveguides is particularly tricky due to the fragility of the thin (180 $\mu m$) fused silica wafer. Handling procedures must be even more stringent because both sides of the substrate should be clean and free from scratches, as the top side hosts the optical waveguides and the emitted light will pass through the bottom side on an inverted microscope. Standard photolithography processes and ion beam etching were used to fabricate tantalum pentoxide strip waveguides on four-inch wafers, and the steps are shown in Fig. S1a and b with additional details provided in the methods section. The wafers were diced into individual chips during the back end processing, which is done in a non-cleanroom environment, and were temporarily bonded with a thick (1 mm) glass wafer (Fig. S2) to protect them from breaking during this process. This provided mechanical stability from the pressure during wafer dicing and polishing, and could be safely removed without damaging the waveguides afterwards, without adding significant cost to the process. Finished chips on both opaque silicon and transparent fused silica substrates are shown in Fig. 2a and b. For the transparent substrate, waveguides of varying widths were examined by scanning electron microscopy (top view and cross section shown in Fig. 2c and d) and showed no visible fabrication defects.

As the roughness of the substrate and the deposited $Ta_2O_5$ has both a strong impact on the waveguide propagation losses and can influence the aberrations when imaging on an inverted microscope, we used an atomic force microscope (AFM) to examine the roughness both before and after deposition (Fig. S3). While the transparent silica substrate was rougher (rms = 0.460 nm) than a standard oxidized wafer (rms = 0.178 nm), both were significantly less rough than a glass coverslip typically used for microscopy (rms=1.5 - 2 nm)[34]. After sputtered deposition of 250 nm of $Ta_2O_5$ onto the substrates, we measured the new rms surface roughness values to be 1.061 nm for the transparent substrate and 0.632 nm for the standard oxidized substrate; these measurements suggest that the roughness of the base substrate will impact the roughness after subsequent deposition steps, and ultimately will affect the final waveguide losses.

The difference between the RI of the waveguide and the surrounding medium dictate the strength of the evanescent field, thus it was important to measure any changes that occurred in the material during the fabrication process. The RI of the $Ta_2O_5$ film deposited on the fused silica substrate was measured in the visible spectrum using a spectroscopic ellipsometer and compared with the same film deposited on an opaque substrate (Table 1, Fig. S4). The slightly higher RI of $Ta_2O_5$ film on the transparent substrate (2.164 at 660 nm) compared to the same film on the opaque silicon substrate (2.152 at 660 nm) may be due to differences in the substrates' surface properties, including their surface roughness. While the magnitude of this index difference is not expected to affect the field strength significantly, the higher RI is actually preferable as it leads to a stronger evanescent field.

The final quality control check before the waveguides could be used for imaging was to measure the propagation losses using scattering analysis[35]. The propagation loss for narrow (5 $\mu$m wide) $Ta_2O_5$ waveguides on the transparent substrate was found to be 6.6 dB/cm at 660 nm, which is notably higher than for similar waveguides on opaque substrates (3.7 dB/cm). This higher loss is mainly related to the $Ta_2O_5$ film quality, which depends on the initial substrate surface quality, and in the future can be reduced by improving the substrate roughness. For wider waveguides (>50 µm width), the propagation losses were on the order of 1 dB/cm; these lower losses are because the modes of the propagating light have less overlap with the roughness of the waveguide sidewalls. As a wider waveguide translates to a larger field of view, all of the c-TIRF imaging reported here used these wider, lower loss waveguides (200 - 500 µm), which are still narrow enough to ensure sufficiently high power density for imaging.

**3. Fluorescence imaging using waveguides on a transparent platform**

As our transparent platform provides imaging capability with both upright and inverted configurations, we have developed an experimental setup that integrates both these configurations (Fig. 3) thereby allowing us to directly compare images taken in both modalities. To ensure one-to-one comparison we have used identical objective lenses (60X 1.2 NA water immersion) and CMOS cameras. In addition, to reduce the impact of photobleaching between measurements in the two different modalities, low power was used for excitation and the measurements were performed consecutively. A well-like sample chamber consisting of an approximately 150 µm thick layer of polydimethylsiloxane (PDMS) was placed on the top of the waveguide surface to hold the sample and the aqueous imaging buffer, and was sealed with a #1.5 thickness coverglass[24].

To allow for a large illuminated FOV we chose to use very wide waveguides, e.g. 200 - 500 µm wide, which will simultaneously guide several modes inside. The interference of these modes creates a wavy, uneven illumination pattern, which can generally be evened out by translating the coupling objective lens with respect to the input facet of the waveguide over time, but the lack of certain modes can lead to stripe-like artefacts as seen in some of the images. However, this uneven illumination pattern can be exploited for fluctuation-based super-resolution imaging, and because of the high RI of the waveguide material the illumination patterns can contain higher spatial frequencies than with conventional free-space optics[2], potentially leading to even higher resolution.

While TIRF microscopy provides illumination of only the thin bottom section of the sample (100 - 200 nm), the emitted fluorescent light can undergo additional scatterings and refraction, causing it to appear out of focus during imaging. This phenomenon is especially pronounced in an upright configuration, as the emitted fluorescence has to travel through an inhomogeneous sample and imaging buffer before reaching the collection objective lens, thus increasing the opportunity for scattering and refraction. In contrast, when imaging in inverted mode the emitted fluorescence only needs to pass through the relatively homogeneous and non-scattering chip to reach the objective lens. To evaluate the impact of the difference in collection pathways, we have collected c-TIRF images of solutions of standard 2 $\mu$m diameter fluorescent beads using both the upright and inverted microscopy setups. In the image of an aggregate of beads taken with the upright setup, shown in Fig. 4a, a diffuse fluorescent signal can be observed which makes it somewhat difficult to identify and to resolve the boundaries between individual beads. In comparison, a c-TIRF image of the same area acquired using the inverted configuration (Fig. 4b) shows better contrast and resolution between beads as compared to the upright configuration, since there is less scattering and refraction of the emitted light. The difference between the two modalities shows more clearly in the large aggregation of beads than it does when there are only two beads, as shown in the upper right corner within Fig. 4. This demonstrates that, for c-TIRF images of beads and other samples with high RI mismatch or scattering potential, an inverted microscopy configuration will yield images with fewer aberrations and better resolution due to the more direct path to the imaging objective. Thus, the enabling of an inverted configuration through development of a transparent chip has a real impact on the image quality compared to opaque chips, which are limited to upright configurations. The transparent chip still maintains the c-TIRF advantage of scalability of the FOV, as demonstrated in Fig. S5, which shows c-TIRF imaging of beads over a FOV of 635 x 635 $\mu m^2$ using a 20X 0.45 NA air objective lens.

We expanded this investigation to include biological samples and chose the relatively thick HeLa cell line (10 - 20 $\mu$m thick) to emphasize the problems associated with scattering. The thin structures of actin networks in cells, here labelled with Atto647N Phalloidin, can provide a distinct, sub-diffraction sized target to quantify the resolution of different imaging techniques in fluorescent microscopy. Fixed HeLa cells were imaged on the same dual configuration microscope as the fluorescent beads, and the averaged c-TIRF images are shown in Fig. 5. While the images from both configurations may initially appear quite similar, there appears to be slightly higher resolution in some areas with the inverted setup (Fig. 5c) than with the upright configuration (Fig. 5b). In particular, the region under the nucleus is less visible with the upright configuration, and this relatively dark region is broader than would be expected if it was purely due to uneven waveguide modes. The blurring effect is likely due to scattering from the relatively thick and dense nucleus, which results in additional aberration and reduced collection of the emitted light. However, imaging with our inverted setup resulted in an overall slightly reduced signal compared to the upright setup, which reduced the improvement in image quality in the thinner, less-scattering parts of the sample. We hypothesize that this is due to minor differences in the optical path length of the chip compared to the coverslip the objective was designed for, which can be compensated for by adjusting the chip thickness in future designs or by increasing the intensity or exposure time when using the current chip design to achieve a similar signal-to-noise ratio. The ability to image in both modes as needed, and the improvement under dense areas of the sample both demonstrate the utility of the transparent c-TIRF platform. Another important characteristic for the c-

TIRF platform is biocompatibility. While the growth of the HeLa cells on these chips already confirms the biocompatibility shown previously for $Ta_2O_5$ waveguides[18], live HeLa cells labelled with SiR-Actin were additionally imaged using the inverted setup as a demonstration (Fig. S6). In the future, however, a stage-top incubation system will be necessary to ensure long-term viability.

Next, we have demonstrated the implementation of a high-NA oil immersion objective lens (60X 1.49 NA) with c-TIRF in the inverted configuration; due to its short working distance it was not possible to image in the upright configuration using this objective. We used immersion oil with an RI of 1.46 (refractive index liquid, Cargille Labs) to match the RI of the fused silica substrate and reduce spherical aberration, with further fine-tuning using the correction collar of the objective lens. Beads imaged using this 1.49 NA oil objective lens (Fig. 4c) are better resolved than with the 1.2 N.A. water immersion objective lens (Fig. 4b), and the smaller depth of focus helps to reject the light scattered and refracted from elsewhere in the bead aggregate. Similarly, images of actin in fixed HeLa cells taken using this 1.49 NA lens on the inverted setup are shown in Fig. S7. Adding the ability to use high NA lenses significantly improves the utility of the c-TIRF platform; while traditional lens-based TIRF systems are locked to this high NA, c-TIRF can be used with a low NA, large FOV lens to screen for interesting events and then be switched over to a high NA lens for enhanced resolution without ever losing the optical sectioning of TIRF.

Finally, we extended the application of the transparent waveguide platform to super-resolution optical microscopy (nanoscopy). We chose to use super-resolution radial fluctuations (SRRF)[25] as it is well-known to provide decent reconstructions at a low signal-to-noise ratio. Here, we take advantage of the fluctuations provided by varying the mode patterns in the waveguide, and instead of averaging them to acquire a standard c-TIRF image, they are processed using the SRRF algorithm in FIJI. The resulting SRRF image (Fig. 6b, d) shows a clear enhancement in resolution compared to the standard c-TIRF image (Fig. 6 a,c). Line profiles (Fig. 6 e, f) drawn across bundles of actin filaments in identical locations in Fig.6 b and d highlight this even further. While the SRRF profile of the first bundle (Fig.6e, purple) shows two clearly resolved actin fibers, the c-TIRF profile (Fig. 6e, green) shows only a slight separation between the same two fibers. For the second actin bundle (Fig. 6f), c-TIRF shows only one large fiber while SRRF has the resolution to distinctly resolve two separate fibers. The overall resolution enhancement is further confirmed with Fourier ring correlation (FRC)[36], which measured the resolution in the SRRF image to be 116 nm (Fig. S8). To the best of our knowledge, this is the first demonstration of super-resolution waveguide-based TIRF on a transparent substrate.

## 4. Discussion and Conclusions

Chip-based TIRF provides significantly improved versatility compared to traditional TIRF techniques by unlocking the restriction of the objective lens and decoupling of the excitation and emission pathways. Additionally, the inherent variation in the illumination pattern in broad waveguides makes the c-TIRF platform a natural partner for fluctuation-based super-resolution techniques, and the high RI of the waveguide enables even smaller patterns to be generated for fluctuations than is possible with free-space optics. As a proof-of-concept, we have shown 116 nm resolution using SRRF on a transparent waveguide, but this can easily be extended to additional super-resolution techniques such as ESI, direct stochastic optical reconstruction microscopy (*d*STORM), or superresolution optical fluctuation imaging (SOFI). These techniques can also be applied to images taken with a lower magnification objective; while the resulting resolution may be above the typical ~250 nm 'super-resolution' limit, it will still be significantly improved compared to both widefield imaging with the same objective or reconstructions of those widefield images using these same fluctuation-based algorithms because the evanescent field from c-TIRF prevents the blurring caused by out-of-focus light.

A major inhibitor of the utility of previous c-TIRF implementations was the opacity of the chip itself, which forced it to be used only on upright microscopes and required significantly complicated sample handling procedures. Additionally, the emitted light needed to travel along a relatively large path through the sample, subjecting it to several sources of distortion before reaching the imaging objective. Here we describe a new c-TIRF platform, and its corresponding fabrication process, which completely eliminates those issues. By fabricating waveguides on a thin, transparent fused silica substrate we have matched these chips to standard glass coverslips in both their imaging and handling properties while still ensuring compatibility with a full range of CMOS mass fabrication processes and the freedom to design a variety of different waveguide shapes and sizes on a mechanically, thermally and optically superior substrate compared to standard glass cover slips. We have shown that these chips can be used on both upright and inverted microscopes on both synthetic and biological samples, and with a variety of imaging objectives, without compromising the excellent optical sectioning of traditional TIRF techniques. Chip-based TIRF can easily accommodate multiple wavelengths, so multicolour imaging and photoswitching are both easily achievable in the future[18].

The use of CMOS-compatible fabrication for these chips means production can be scaled up for eventual mass production at a low cost, allowing them to be used in many more labs and environments than traditional TIRF. Because of the decoupling of the emission pathway from the TIRF excitation, these chips could be combined with more durable cameras and collection optics, enabling them to be directly used in harsher environments, such as inside a humid incubator, under water, or at extremely high temperatures. Alternatively, because the fabrication is CMOS-compatible, chips could conceivably be designed to contain the waveguide, integrated filters, camera, and an LED light source as a single, integrated CMOS-based device. Additionally, as waveguide technology allows light to be sent along an innumerable variety of complex paths not possible in free-space, we see natural extensions for future designs to facilitate e.g. FRAP, photoactivation, or phototaxis along specific patterns. Our aim was to make TIRF microscopy significantly more user-friendly and accessible, and as this promising technology is increasingly adopted we anticipate seeing even more applications in fields where TIRF was previously impractical.

**Acknowledgements**

The authors would like to express their appreciation to Prof. James Wilkinson (University of Southampton) for discussions on the waveguide platform. We thank Sebastian Alberti (UiT) for help with the AFM experiments and Neil Sessions (University of Southampton) for fabrication equipment and SEM imaging. The authors would like to thank Øystein Ivar Helle and David Andre Coucheron for their valuable suggestions. The authors would like to thank Luis Enrique Villegas Hernandez (UiT) for helping in the schematic drawings. This research was funded by the European Research Council Proof of Concept project (H2020, ERC PoC 789817), Research Council of Norway BIOTEK 2021 (Nano-Path, Project No. 285571) and "The Future Photonics Hub" (UK EPSRC grant EP/N00762X/1).

**Author contributions**

B.S.A., S.M.G. and A.P. conceived this project. S.M.G. and B.S.A. supervised and provided funding for the project. A.P. characterized the materials properties and fabricated the waveguide chip. B.S.A., F.T.D., J.C.T. and A.P. contributed to the chip design. J.C.T. assisted with the SEM and AFM analysis. F.T.D. and N.J. characterized the waveguide losses. F.T.D., A.A. and N.J. built the imaging setup and performed the experiments. F.T.D., D.L.W. and N.J. analysed the imaging data. D.L.W. and V.D. prepared biological samples. The manuscript was mainly written by A.P., F.T.D. and D.L.W. with comments from all co-authors.

**Competing interests**

B.S.A. has applied for two patents for chip-based optical nanoscopy. B.S.A. is a co-founder of the company Chip NanoImaging AS, which commercializes on-chip super-resolution microscopy systems.

**Table 1:** Summary of material parameters of the transparent substrate and opaque substrate with and without 250 nm thick $Ta_2O_5$ film on top. Graphs and images for determining the roughness and refractive index of the substrate and film is shown in supplementary Figure S1.

| Substrate | Roughness (rms) | Roughness after $Ta_2O_5$ deposition (rms) | Refractive index by ellipsometry (at 660 nm) | Propagation loss (5 µm wide and 250 nm strip waveguide) |
|---|---|---|---|---|
| **Transparent** Fused Silica (180 µm) | 0.460 nm | 1.061 nm | 2.164 | 6.4 dB/cm |
| **Opaque** Silicon substrate (1 mm) with $SiO_2$ cladding layer (2.5 µm) | 0.178 nm | 0.632 nm | 2.152 | 3.7 dB/cm |

# Figures

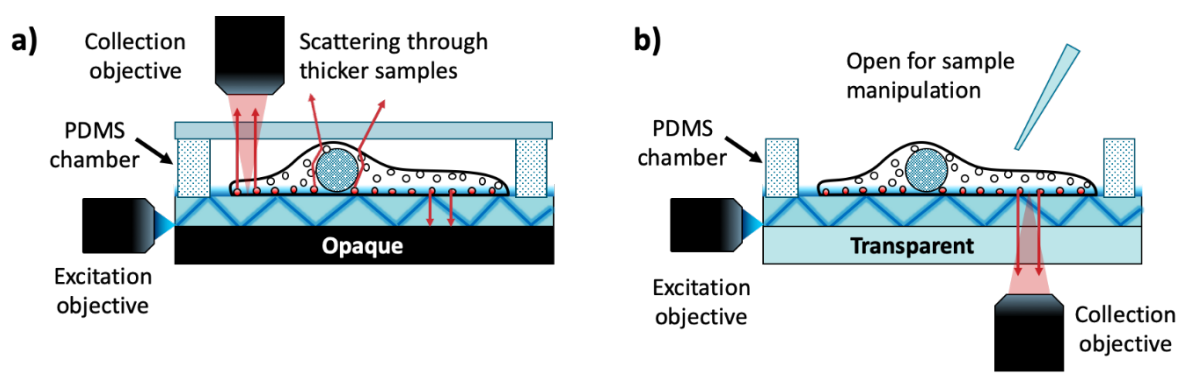

**Figure 1: Schematic illustrating the differences between light paths in opaque versus transparent c-TIRF.** Light is coupled into the waveguide from an excitation objective lens, shown on the left side of the chip. The coupled light is reflected between the interfaces with the substrate (either **(a)** opaque or **(b)** transparent) and the sample, keeping the light totally within the waveguide and generating an evanescent field to illuminate the bottom layer of the sample (shown as a blue line). **(a)** Using a waveguide on an opaque substrate, emitted fluorescence (red arrows) must pass through the bulk of the sample, potentially experiencing scattering or other distortions, before it can be collected by an objective located above the sample in an upright configuration. With a transparent substrate, emitted light can be collected by either an upright configuration as shown in **(a)** or by an inverted microscope as shown in **(b)**. With the inverted setup, the light only needs to pass through the relatively homogeneous chip rather than a distorting sample. Additionally, the area above the sample can be left open to allow for the addition of reagents or for micromanipulation during imaging. Figure is not drawn to scale for the purpose of clarity.

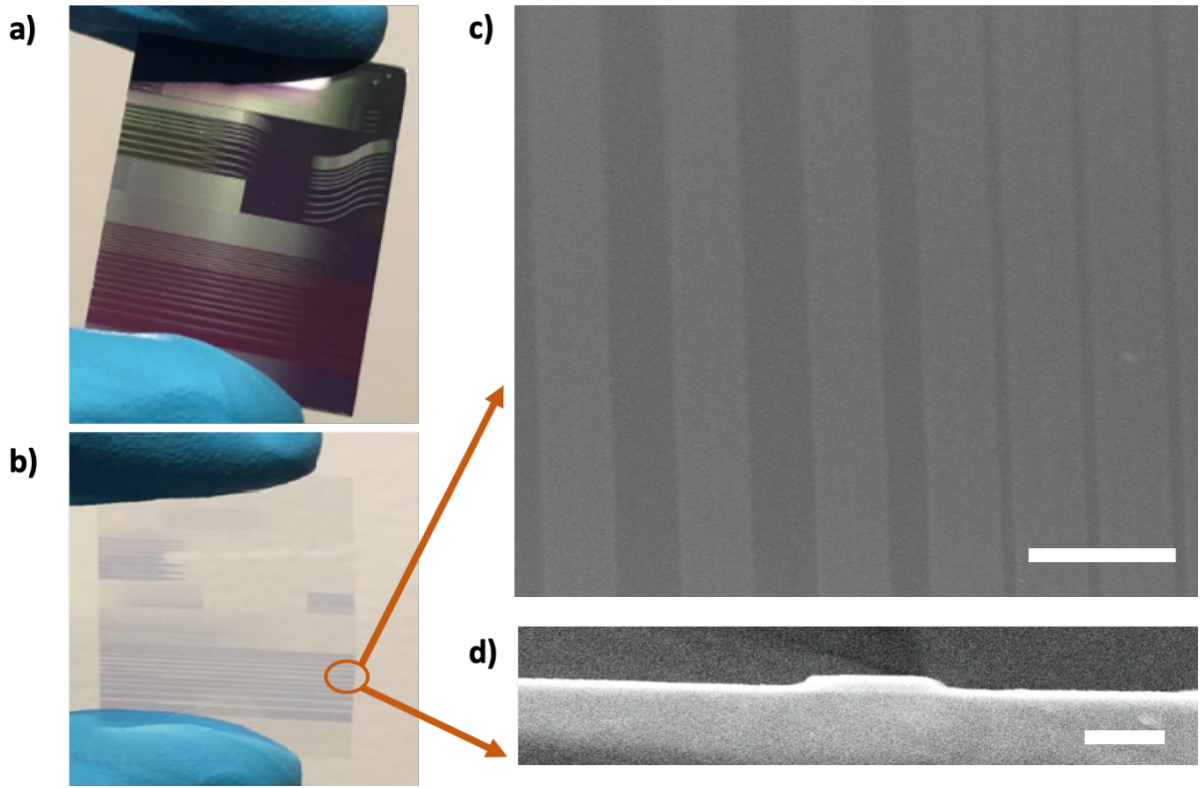

**Figure 2: Fabricated waveguide chips on both transparent and opaque substrates. (a-b)** Photographs of the fabricated waveguide chips after back end processing on either **(a)** an opaque substrate or **(b)** a transparent substrate. **(c-d)** SEM images showing the clean fabrication of a waveguide chip on the transparent substrate from **(b)** a top view, with a scale bar of 10 µm and where the waveguides appear as dark lines and **(d)** a cross-section of the waveguide input facet with a scale bar of 1 µm. The width and height of the fabricated waveguide shown in **(d)** are 1500 nm and 250 nm, respectively, and the sidewall angle is ~85°.

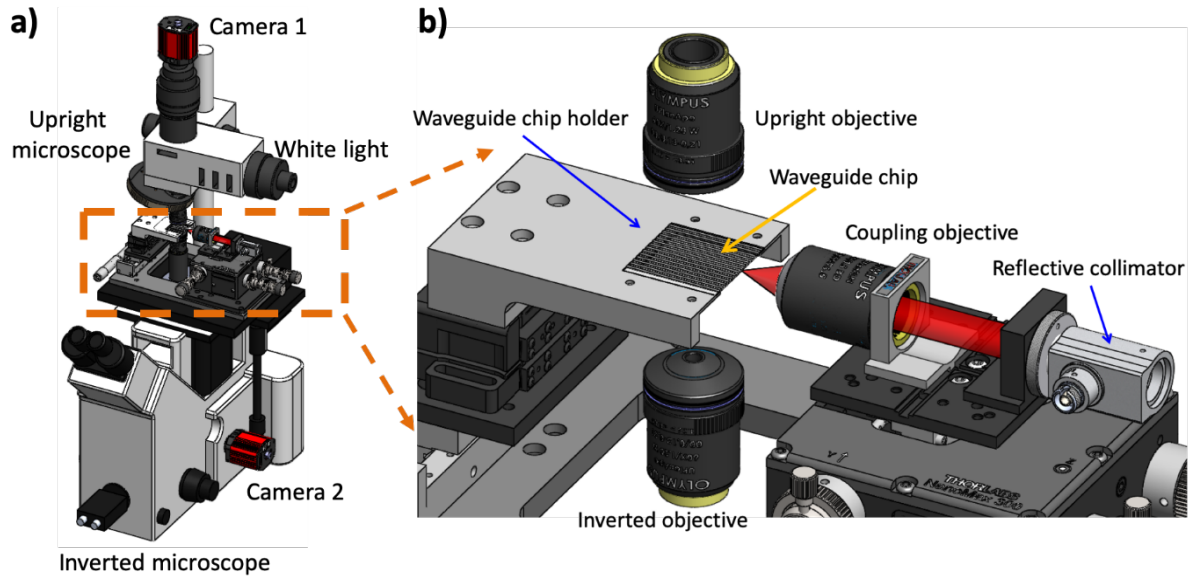

**Figure 3: 3D view of the dual-configuration microscope used for c-TIRF imaging. (a)** 3D rendering of the integrated c-TIRF microscope, where the upright microscope is above the waveguide chip and the inverted microscope is below the waveguide chip. This integrated system enables direct comparison between imaging modalities on the same sample. **(b)** Closeup of the waveguide chip with coupling and imaging objectives. The waveguide chip holder can be rotated in X, Y, and Z to ensure it is aligned flat with respect to the imaging plane, thus avoiding uneven focus across a field of view. It also has a long travel translation stage along the Y-axis for switching from one waveguide to another waveguide on the same chip. The coupling objective and collimator are on an X-Y-Z nanometric stage separate from the main microscope to facilitate free-space coupling of a laser into the transparent waveguide chip.

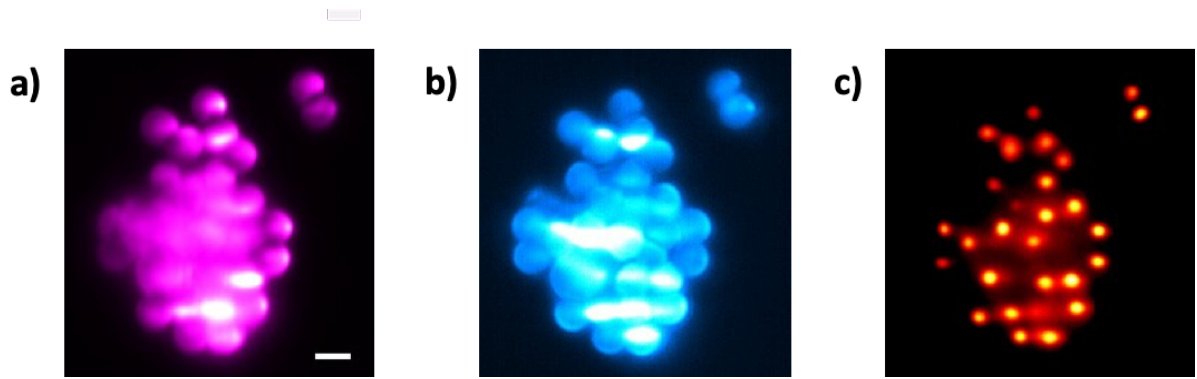

**Figure 4: Comparison of different c-TIRF imaging setups using 2 μm diameter fluorescent beads.** A cluster of beads was imaged using a 60X 1.2 NA water objective lens on a transparent waveguide on **(a)** the upright configuration and **(b)** the inverted configuration, as well as with **(c)** a 60X 1.49 oil lens on the inverted configuration. The c-TIRF image with the inverted configuration in **(b)** shows a clearer separation between beads than compared to the upright configuration in **(a)**, especially in the more dense regions of the bead aggregate. This is due to the strong scattering and refraction of the emitted light as it has to pass through an inhomogeneous sample before reaching the collection objective in the upright configuration. The beads imaged using the 1.49 NA oil objective lens **(c)** are better resolved than with **(b)** the 1.2 NA water immersion objective lens; this is a benefit from having a higher NA and a smaller depth of focus, which helps in axial resolution. All three images were acquired on the same waveguide in the same region and consecutively. In all cases, 1500 frames were averaged and each of these frames was acquired with 50 ms exposure time. Scale bar is 2 μm.

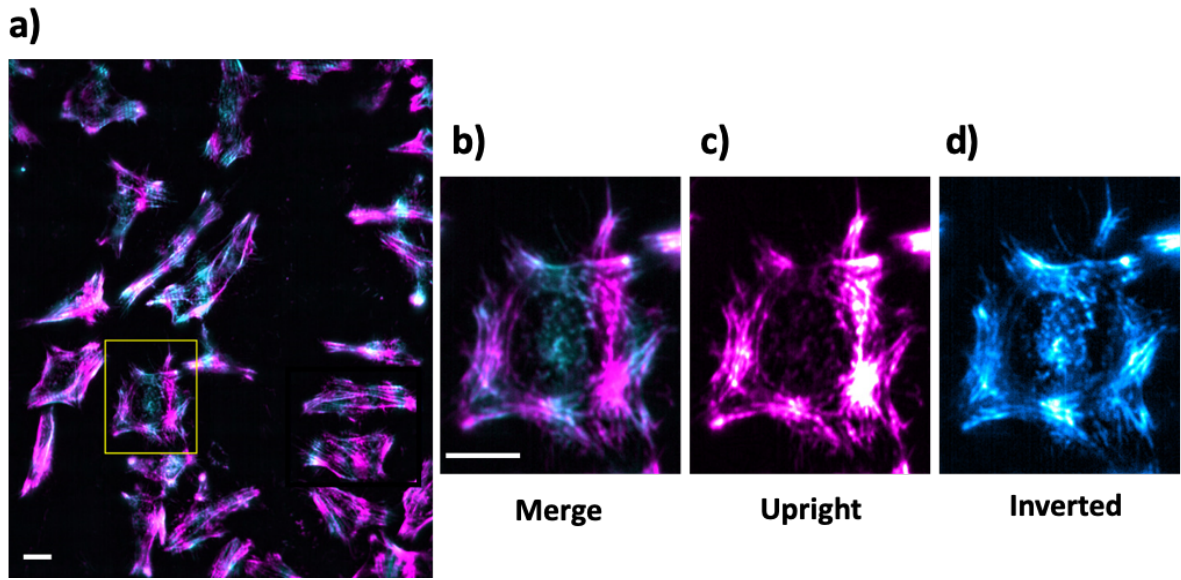

**Figure 5: Comparison of inverted versus upright performance of c-TIRF for imaging actin in fixed HeLa cells.** Cells were imaged on a transparent waveguide using both the **(c, magenta)** upright and **(d, cyan)** inverted configurations, and the two images overlaid as shown in **(a)**; the boxed region in **(a)** is enlarged in **(b-d)**. With the inverted setup **(d)**, more structure is visible under the nucleus than with the upright setup **(c)**; this is due to scattering and refraction of the emitted light as it passes through the bulk of an inhomogeneous sample while traveling to the upright objective, whereas the light only needs to pass through the relatively homogeneous substrate to reach the inverted objective. Both imaging modalities show relatively dark, approximately vertical regions which are due to the mode patterns in the waveguide, but the dark region under the nucleus in **(c)** is broader than can be explained simply by these mode patterns. For all images shown here, 60X 1.2 NA water immersion objectives were used, and 1500 frames each with 50 ms acquisition times were averaged to compensate for the mode patterns. Scale bars are 10 μm.

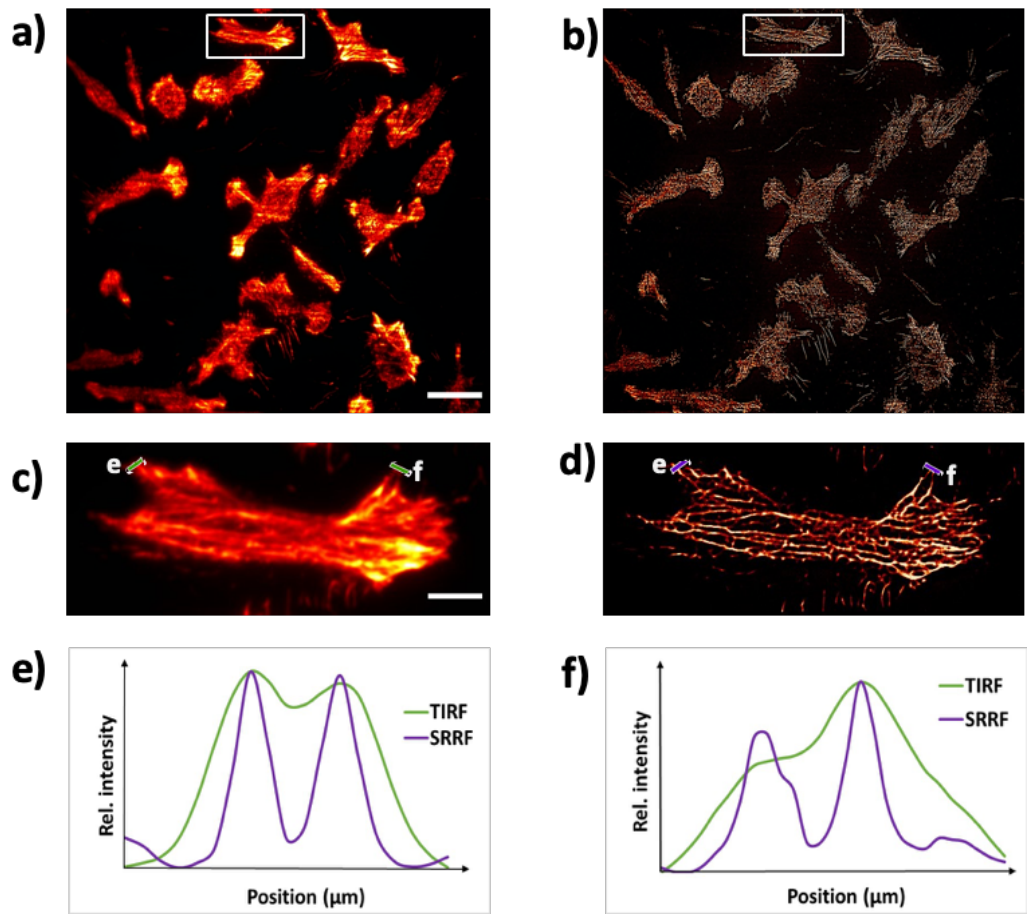

**Figure 6: c-TIRF and super-resolved image of actin in fixed HeLa cells on a transparent waveguide.** Here, a series of images were acquired using a water immersion objective lens (60X, 1.2 NA) on an inverted setup. **(a)** shows an averaged c-TIRF image of actin in fixed HeLa cells with the corresponding super-resolved SRRF reconstruction in **(b)**. **(c – d)** shows the zoomed-in region of the boxes outlined in **(a – b)**. Intensity profiles of the green and purple lines drawn across adjacent actin filaments in **(c – d)** are plotted in **(e-f)**, and show much clearer separation between actin fibers in the SRRF image **(purple lines)** compared to c-TIRF **(green lines)**. By using Fourier ring correlation (FRC) on the SRRF image an optical resolution of 116 nm was measured (Fig S7). For averaged c-TIRF and SRRF images, 1500 and 500 frames were used respectively and each of these frames was acquired with 50 ms exposure time. The scale bar in **(a – c)** is 25 $\mu$m and in **(c – d)** is 6 $\mu$m.

## Methods

**Waveguide fabrication front end process (Fig. S1a)**

250 nm thick $Ta_2O_5$ film was deposited using RF magnetron sputtering directly onto 4" diameter wafers of 180 $\mu$m thick transparent fused silica or of 1 mm thick silicon (with 2.5 $\mu$m $SiO_2$). The base pressure of the deposition chamber was kept below 1 x $10^{-6}$ Torr with $Ar:O_2$ flow rates of 20 standard cubic centimetre per minute(sccm) : 5 sccm. A substrate temperature of 200°C was maintained throughout the deposition time. $Ta_2O_5$ films were deposited with a high deposition rate of 2.2 nm/min as determined using a stylus profiler (KLA Tencor P-16+ model). Photolithography was used to create a photoresist mask for further dry etching to fabricate channel waveguides. First, a 1.3 $\mu$m thick layer of positive resist (Shipley S1813) was spin-coated on top of the 250 nm $Ta_2O_5$ film and then prebaked at 90°C in an oven for 30 minutes. Second, the prebaked photoresist was exposed under a hard chrome mask with waveguide patterns using a mask aligner (MA6, Carl Suss). The time needed to properly expose the photoresist differs depending on the substrate used (based on whether it is reflective, transmissive or absorptive), therefore the exposure time was optimised separately for the transparent substrate (5.8 seconds at 20 mW/$cm^2$) and opaque substrate (5.6 seconds at 20 mW/$cm^2$). Finally, the exposed patterns were developed using MF-319 developer for 58 seconds.

The dry etching was performed in an ion beam system (Ionfab 300+, Oxford Instruments) using Argon gas with a flow rate of 6 sccm and Trifluoromethane ($CHF_3$) gas with a flow rate of 12 sccm to fully etch the $Ta_2O_5$ waveguides. The process beam voltage (500 V), beam current (100 mA), RF power (500 W) and substrate temperature (15°C) were kept constant throughout etching process. In the ion beam milling process, the substrate was placed at an angle of 45° with respect to the incident ion beam to achieve low sidewall roughness. The processed wafer was subjected to plasma-ashing with oxygen gas for 20 minutes to remove the photoresist. After that, the wafers were placed in a 3-zone tube furnace for annealing at 600°C (ramp rate of 3°C/min to 525°C, ramp rate of 2°C/min from 525°C to 600°C) in an oxygen environment (2 liter/min) for 3 hrs to reduce the stress and repair the oxygen deficiency created in $Ta_2O_5$ during the fabrication process[37].

**Waveguide fabrication back end end process (Fig. S1b)**

Photoresist (MICROPOSIT®S1813) was spin-coated on top of the processed wafer on both sides of the substrate to safeguard waveguides from scratching and contamination during the back end process. Thin fused silica substrate was temporarily bonded with a thick (1 mm) glass wafer with a 4" diameter to ensure that the thin transparent wafer would not break during the back end processing. As shown in Fig. S2a, the thin transparent waveguide wafer was placed on the thick circular glass carrier substrate using quartz wax from Logitech (part no. 0CON-200, Melting point 66 – 69°C) and heated to 80°C for its fast melting. Constant pressure was applied after melting the wax to make sure no air bubble remained trapped and that spreading of the wax was uniform throughout the thin transparent wafer as shown in Fig. S2b. After that, the transparent wafer with carrier glass was diced with a diamond blade under water cooling (Isomet 5000, Buehler) into small chips.

Instead of dicing for the opaque substrate, cleaving along the crystal lattice was performed to dice the chips.

The lapping and polishing processes for the transparent and opaque chips were carried out on a Logitech LP50 system. The initial lapping step utilized a 9 μm alumina solution from 30 minutes to one hour, until the waveguide end facets of the diced chips were fully visible and flat. The second lapping step utilized a 3 μm alumina solution and was performed for 30 minutes to reduce the surface roughness to below 3 μm. In both cases, the sample was placed against a rotating metal plate. The alumina lapping solution with 3 μm particles offered sufficient removal rates of 5 μm min$^{-1}$. After lapping, the samples were polished on a different LP50 system using a diamond slurry. The surface roughness value for diced waveguide input facets was reduced from a few microns to tens of nanometers after polishing.

**AFM Measurement**

Roughness of the substrates and deposited $Ta_2O_5$ thin films was measured with an AFM (Cypher, Asylum Research) in amplitude modulation mode by scanning an area of 2 $\mu$m by 2 $\mu$m. A standard cantilever (Olympus AC160TS, k ≈ 30 N m$^{-1}$, Q ≈ 300, f(1)0 ≈ 300 kHz), was used for all experiments. A scanning rate of 1 Hz and an oscillation amplitude of ~7 nm was employed for the roughness measurements. For reconstructing the tip-sample interaction force, the amplitude and phase curves were recorded versus the tip-sample separation distance, and Sader–Jarvis–Katan formalism was used to transform the cantilever dynamics into the force versus distance curves[38].

**Ellipsometry Measurement**

The ellipsometric spectra of the 250 nm thick $Ta_2O_5$ film on the transparent fused silica and opaque silicon substrates were collected using a J.A. Woollam VASE ellipsometer in the 400–800 nm spectral range. The measurements were performed at an angle of incidence of 55° and analysed using the CompleteEASE ellipsometry data analysis program. They were fitted using the Tauc-Lornetz method which is suitable for amorphous semiconductor materials[39].

**Microscopy setup for c-TIRF**

Fig. 3 shows the 3D schematic of the c-TIRF microscopy system developed to image in both inverted and upright directions. The microscope base is a Leica DMIRB, and the sample stage area (Fig. 3b) has been modified to work with waveguide chips. A chip is held on an aluminium holder with an open bottom, which simultaneously provides access to the upright objective above, the inverted objective below, and the coupling objective on the side. Laser excitation light (iChrome MLE, Toptica Photonics, 100 mW, 561 or 640 nm; and Cobolt Flamenco, 500 mW, 660 nm) is coupled into a single mode fiber (P1-460Y-FC-1, Thorlabs) and collimated using a reflective collimator (RC04FC-P01, Thorlabs) before being focused by a coupling objective (Olympus 50X 0.5NA) onto a side facet of the waveguide. The positioning of the laser focus on the input facet can be finely adjusted using a X-Y-Z nanometer precision stage (Nanomax 300, Thorlabs) to ensure the majority of the light is coupled into the waveguide, and can be further adjusted to change the mode patterns inside the waveguide using a piezo controller (MDT693B Open-Loop Piezo Controller, Thorlabs). The waveguide chip holder is attached to a single-axis, long range linear translation stage (7T173-20-50, Standa) which is used to switch the coupling of the light between multiple waveguides on the same chip, as well as to a 6-axis stage (9031-M, Newport) which can be used to eliminate tilt on the waveguide chip relative to the imaging plane, thus ensuring the entire FOV will remain in focus. Fluorescent light emitted by the sample on the waveguide is captured by either the upright or inverted objective and is filtered through a combination of longpass and bandpass filters (BLP02-561R-25, Stock#67-034;BLP01-664R-25, FF01-692/40-25, respectively, SEMROCK) to block the excitation light before being imaged onto a CMOS camera (ORCA-Flash4.0 v2, Hamamatsu). Identical objectives, cameras, and filters are used in both the upright and inverted pathways to ensure an equal comparison between imaging modalities. A summary of the imaging settings used for all of the figures is included in Table S1.

**Propagation loss of the $Ta_2O_5$ waveguides**

Propagation losses were analysed using the scattering analysis method[35]. 660 nm laser light was coupled into the waveguide and propagation losses were measured by acquiring images of light scattered from the waveguide surface using a microscope setup previously described[35]. To avoid saturated pixels, the exposure time for the microscope's CCD camera was adjusted. The images were processed with a MATLAB program: an average value for each image was used to find the intensity of propagating light as a function of the waveguide length, and then the result was plotted on a log scale with a linear fit to estimate the propagation loss (dB/cm).

**Sample preparation**

2 μm diameter, uniformly fluorescent microspheres with an RI of 1.59 and excitation/emission wavelengths of 540/600 nm (FC05F, Bang Laboratories) were used. The stock solution was diluted 1:100 in water and 500 μl of the solution was added into the PDMS well chamber on the transparent waveguide.

HeLa cells were grown in minimum essential medium (MEM) supplemented with 10% fetal bovine serum and 1% penicillin/streptomycin, in a standard humidified incubator at 37 °C with 5% $CO_2$. Cells were seeded into the PDMS well chambers on transparent waveguides (described further in Section 3) 1-2 days before imaging. Cells used for live-cell imaging were incubated with 1 μM SiR-Actin and 10 μM verapamil (Spirochrome) in culture medium for 1 hour before washing. The culture medium was aspirated and replaced with pre-warmed Live Cell Imaging Solution (Invitrogen) before imaging. For phalloidin labeling, HeLa cells were similarly seeded within PDMS chambers, and after 1-2 days of growth fixed for ~20 minutes using 4% paraformaldehyde in PBS. Cells were then washed in PBS followed by incubation with 0.1% Triton X-100 in PBS for 4 minutes, and washed 3 times in PBS for a few minutes each time. Cells were then incubated with Atto-647N phalloidin (1:33 for widefield c-TIRF, 1:100 for SRRF imaging) in PBS for 90 minutes.

**Image processing**

All microscopy images were processed using FIJI[40,41]. Image sequences with fluctuating modes were first averaged over time, followed by processing steps performed on the averaged image. Background subtraction with a rolling ball radius of 50 pixels was performed for c-TIRF images of cells. Slight differences in rotation and magnification between the cameras for the upright versus inverted camera were compensated for using the Align Image by line ROI plugin for the images of fixed cells; this allowed for direct comparison by merging/overlaying channels. Images shown in this paper have been linearly adjusted for brightness and contrast. Single-channel images use a 'hot' style lookup table for improved contrast.

The super-resolved images were processed with the SRRF plugin for FIJI[25]. Five hundred (500) frames were used to process the SRRF images, each image having 50 ms acquisition time. A ring radius of 0.5 (default) and radiality magnification of 15 were used.

Supplementary Information

# A transparent waveguide chip for versatile TIRF-based microscopy and nanoscopy

Anish Priyadarshi[‡a,b*], Firehun Tsige Dullo[‡a], Deanna L. Wolfson[a], Azeem Ahmad[a], Nikhil Jayakumar[a], Vishesh Dubey[a], Jean-Claude Tinguely[a], Balpreet Singh Ahluwalia[a*] and Ganapathy Senthil Murugan[b*]

[a] Department of Physics and Technology, UiT The Arctic University of Norway, 9037 Tromsø, Norway

[b] Optoelectronics Research Centre, University of Southampton, Southampton SO17 1BJ, United Kingdom

‡ contributed equally to this work.

**Corresponding emails:** anish.priyadarshi@uit.no, balpreet.singh.ahluwalia@uit.no , smg@orc.soton.ac.uk

**Table S1:** Overview of the imaging acquisition parameters

| Figure number | Sample | Excitation wavelength (nm) | Emission wavelength (nm) | Acquisition time (number of frames x exposure time) | Objective lens |
|---|---|---|---|---|---|
| Fig. 4 | Fluorescent beads | 561* | 590 | 1500 x 50 ms | 60X 1.2 NA Water 60X 1.49 NA Oil |
| Fig. 5 | Actin in fixed HeLa cells | 640* | 690 | 1500 x 50 ms | 60X 1.2 NA Water |
| Fig. 6 | Actin in fixed HeLa cells | 660** | 690 | 1500 x 50 ms | 60X 1.2 NA Water |
| Fig.S5 | Fluorescent beads | 561* | 590 | 1500 x 50 ms | 20X 0.45 NA Air |
| Fig.S6 | SiR-Actin in live HeLa cells | 640* | 690 | 2500 x 10ms | 60X 1.2 NA Water |
| Fig.S7 | Actin in fixed HeLa cells | 640* | 690 | 1500 x 50 ms | 60X 1.49 NA Oil |

* iChrome MLE, Toptica Photonics, 100mW

**Cobolt Flamenco, 500 mW

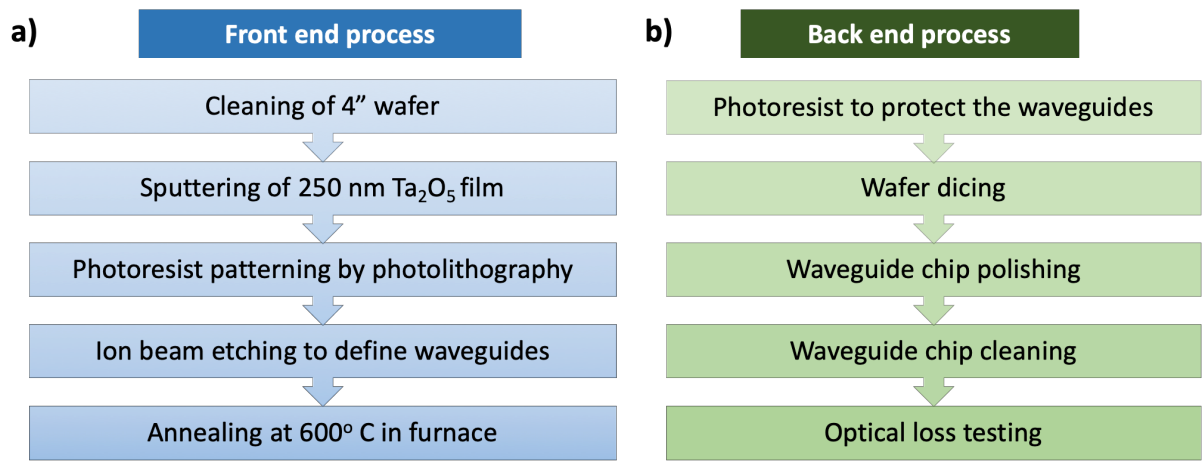

**Figure S1: Fabrication steps to produce Ta$_2$O$_5$ waveguides on transparent and opaque (silicon) substrates. (a)** Front end and **(b)** back end process steps for waveguide chip fabrication are the same for both types of substrates, except for a temporary support of the thin transparent wafer during back end processing; this step is described further in Figure S2.

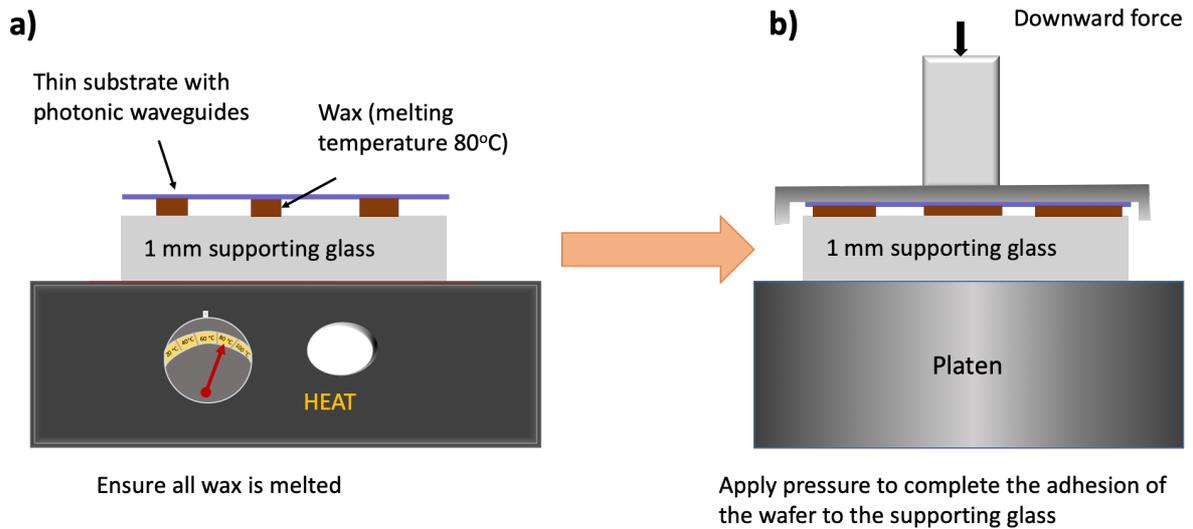

**Figure S2: Steps for fastening the thin transparent wafers to the 1 mm glass support for back end processing. (a)** Wax was heated to 80°C on top of the 1 mm thick supporting/carrier glass, and the transparent waveguide wafer was placed on top of the wax after it melted. **(b)** A downward force was applied using a flat steel plate and additional weight was added on top to eliminate trapped air and ensure a thin and uniform layer of wax. The thick glass provided structural support so that the fragile, ultrathin (180 μm) wafer was protected from breaking during the harsh back end steps.

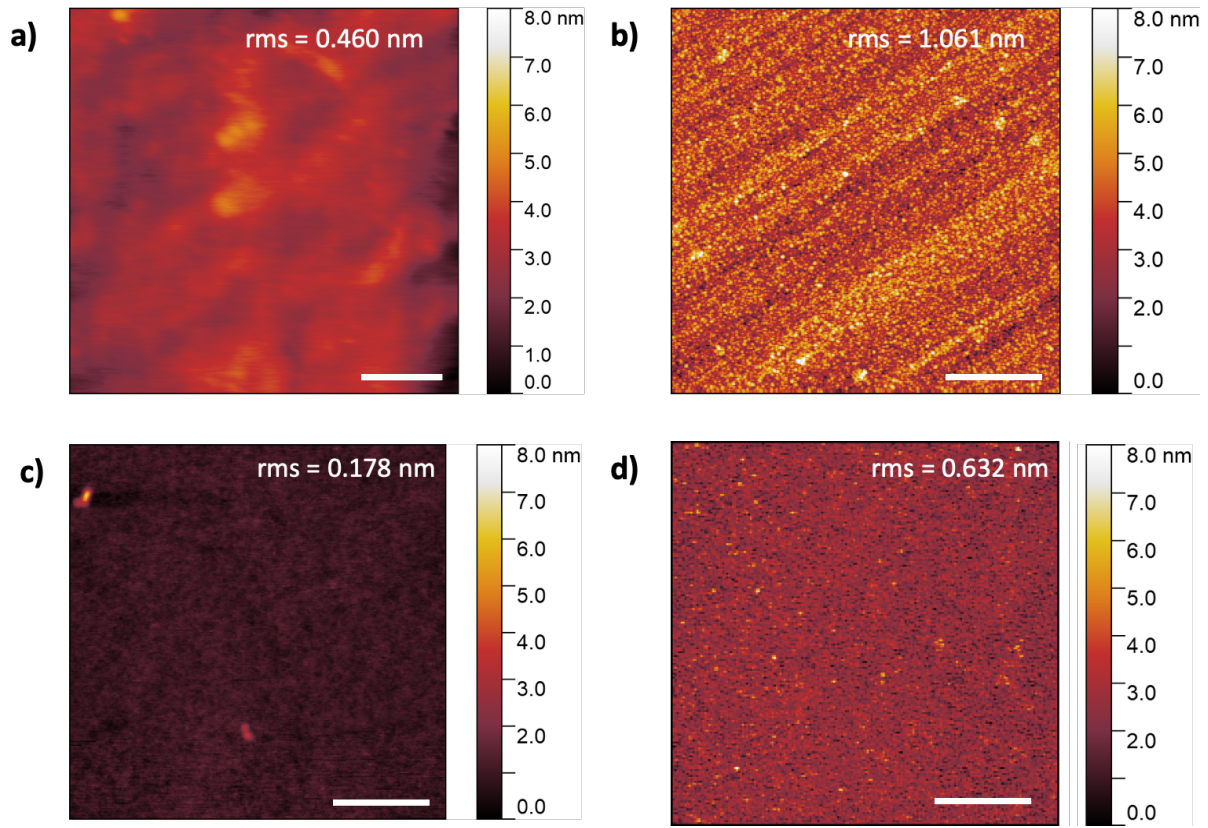

**Figure S3: Surface roughness of different substrates with and without Ta$_2$O$_5$ film measured by AFM.** Topography images show the surface roughness of **(a)** the transparent substrate, **(b)** the transparent substrate with sputter-deposited Ta$_2$O$_5$ film, **(c)** the opaque substrate, and **(d)** the opaque substrate with sputter-deposited Ta$_2$O$_5$ film. The rms roughness values extracted from these images are listed in Table 1 of the main manuscript as well. Scale bar is 500 nm.

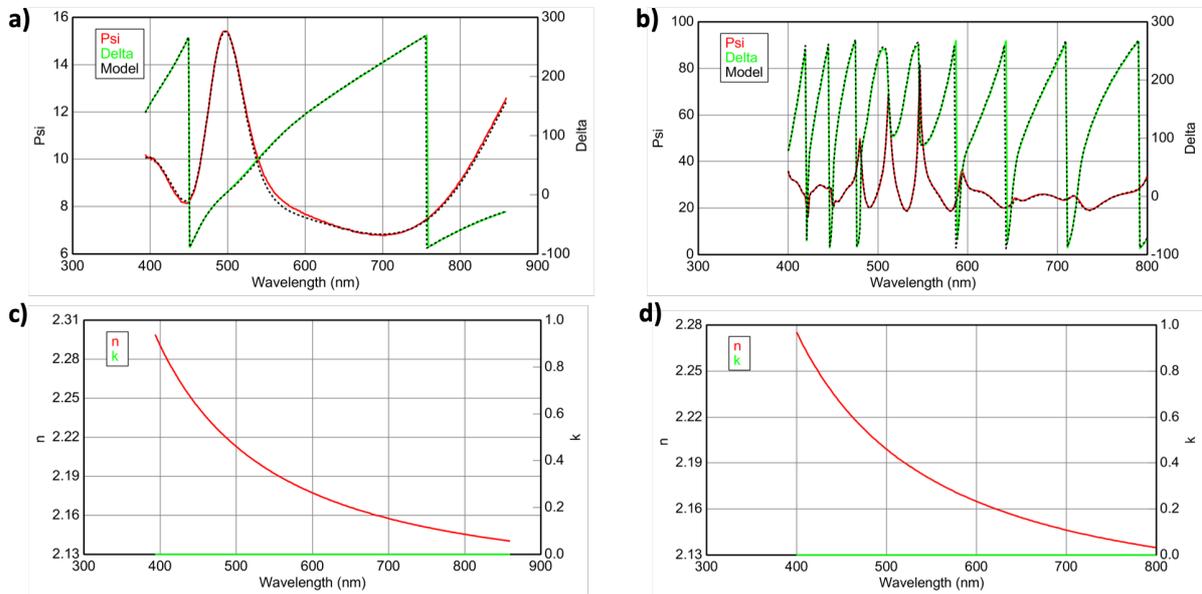

**Figure S4: Spectroscopic ellipsometry (SE) measurements of optical constants for the 250 nm thick Ta₂O₅ film on transparent and opaque substrates**. **(a-b)** shows the experimental and modelled ellipsometric parameters (psi and delta) as a function of wavelength (400 nm – 800 nm) for **(a)** transparent and **(b)** opaque substrates, respectively, as measured at a 55° incident angle. Experimental ellipsometric data were fitted using the Tauc-Lornetz method, which is suitable for amorphous semiconductor materials, to find the RI (n) and extinction coefficient (k). The calculated optical constants (n and k) for the Ta₂O₅ film on **(c)** transparent and **(d)** opaque substrates are shown plotted as a function of wavelength. The RI values extracted from this data at a wavelength of 660 nm are listed in Table 1 of the main manuscript**.**

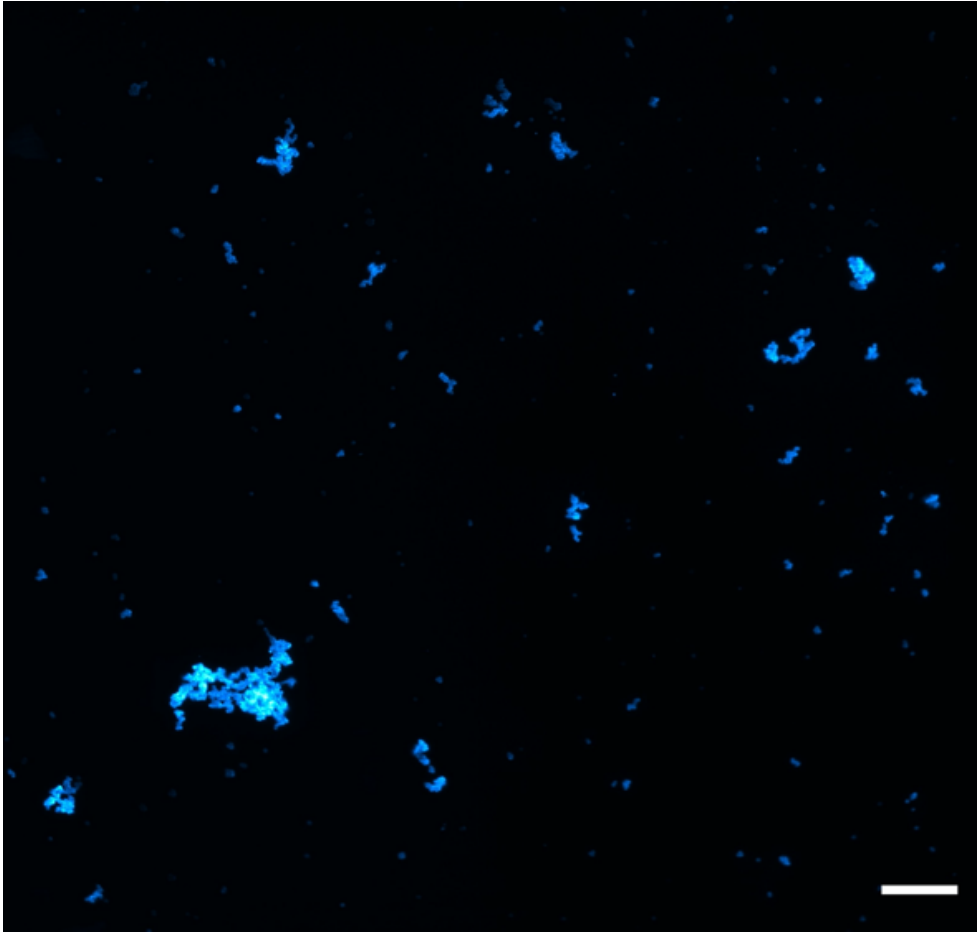

**Figure S5: Large FOV c-TIRF image of 2 μm diameter fluorescent beads.** Bead aggregates similar to those shown in Fig. 4 were also imaged using transparent c-TIRF in an inverted configuration, here with a 20X 0.45 NA air objective to capture a much larger FOV. The FOV shown here is 635 x 635 μm$^2$; this is approximately 100X larger than what is possible using traditional objective lens-based TIRF, but without any compromises to the axial resolution inherent to TIRF. Scale bar is 50 μm.

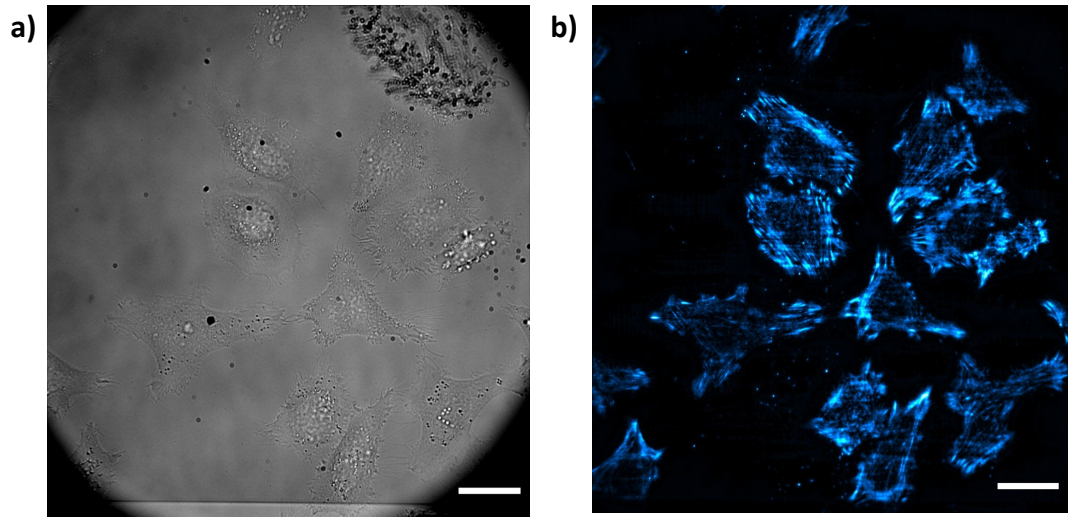

**Figure S6: Brightfield and c-TIRF images of actin in living Hela Cells.** HeLa cells on a transparent waveguide were labelled with SiR-Actin and imaged using **(a)** brightfield, and **(b)** c-TIRF on an inverted setup, both with a 60X 1.2 NA water immersion objective lens. The cells appeared healthy during the short timeframe they were imaged, but a stage-top incubator would be necessary for longer-term imaging. For the c-TIRF image, 2500 frames were used and each of these frames was acquired for 10 ms. Scale bar is 10 μm.

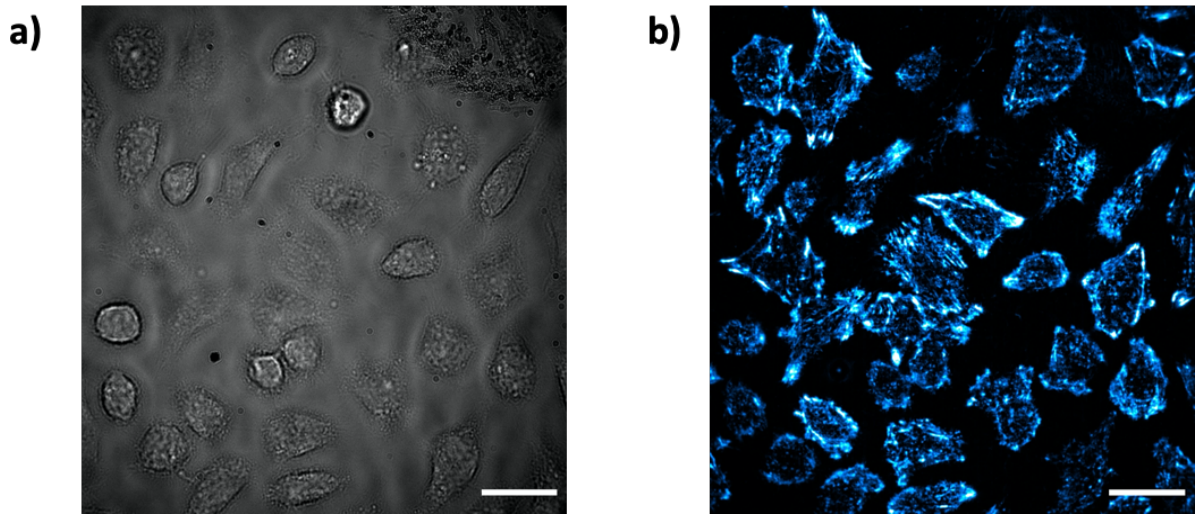

**Figure S7: Brightfield and c-TIRF images of actin in fixed HeLa cells taken using a high NA objective lens on a transparent waveguide.** HeLa cells were fixed and imaged in **(a)** bright field and **(b)** with c-TIRF using a 60X 1.49 NA oil immersion objective in an inverted configuration. A series of images were acquired using an oil immersion objective lens (60X 1.49 NA) and inverted setup. For the c-TIRF image, 1500 frames were used and each of these frames was acquired for 50 ms. Scale bar is 10 μm.

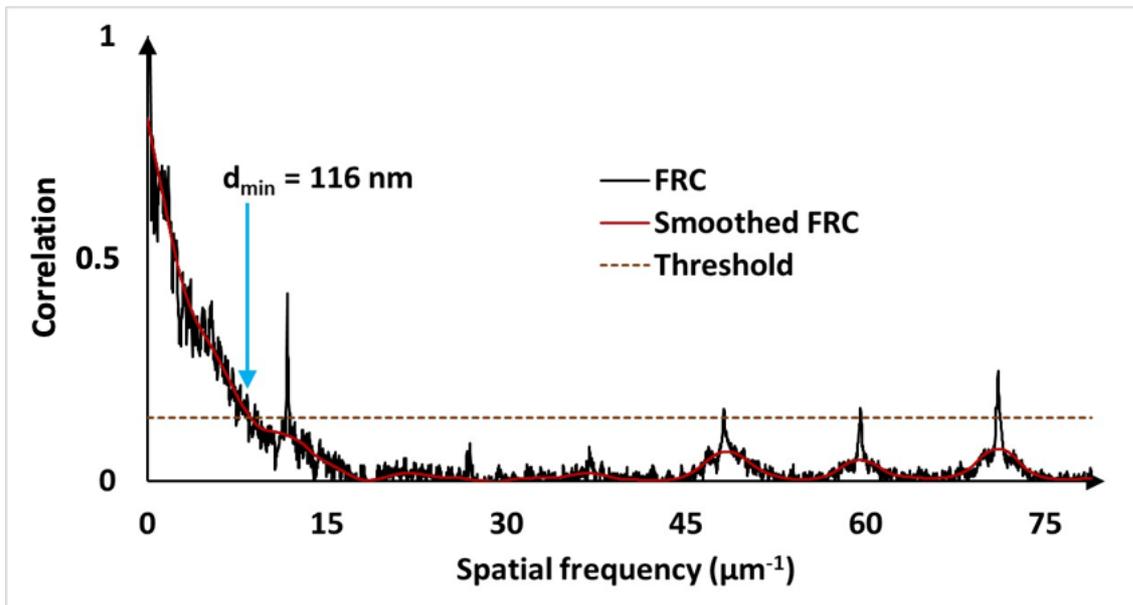

**Figure S8: Fourier ring correlation shows a SRRF resolution of 116 nm.** The SRRF image shown in Fig. 6 was analysed using the FRC plugin in FIJI, and the resulting correlation curve is shown here. Resolution is extracted from the spatial frequency at which the correlation value drops below a standard threshold of 1/7 (dashed line), which here corresponds to a lateral resolution of 116 nm (blue arrow).